 \documentclass[preprint]{aastex}

\shorttitle{Rajagopal et al.}
\shortauthors{The confusion limit for SIM}

\begin{document}


\title{The confusion limit on astrometry with SIM}


\author{Jayadev Rajagopal, Torsten B\"oker\altaffilmark{1} and Ronald J.Allen }
\affil{Space Telescope Science Institute, Baltimore, MD 21218}
\email{jayadev, boeker, rjallen@stsci.edu}


\altaffiltext{1}{Affiliated with the Astrophysics Division, Space Science Department,
European Space Agency}


\begin{abstract}
An important requirement for the Space Interferometry Mission (SIM)
is to carry out precision astrometry in crowded fields. This capability
is crucial, for example, to accurately measure proper motions of bright stars 
in nearby galaxies. From such measurements, one can obtain distance estimates, explore the dynamics 
of these systems and measure the mass of the Milky Way itself. In this paper we investigate 
errors introduced by confusion,
i.e. the presence of objects other than the targeted star in the SIM
field of view (FOV). Using existing HST images
of fields in M31, the LMC and the Galaxy we 
simulate the background within the SIM FOV and estimate the errors in the 
measured position of the target star. Our simulations account for the error contribution from photon statistics. 
We also study the effects of pointing imperfections 
when a field is revisited which result in errors in the
measured proper motion. 
We use the simulations to explore
the measurement accuracy of several SIM key programs which will require crowded
field astrometry.  In M31, the error in the absolute position of the targets could 
be significant for all but the brightest targets. 
Our results also indicate that in the case of 
the brightest targets in M31, and for all likely target magnitudes in the other
cases, confusion-induced proper motion errors are well within the 
SIM requirements. However, targets which vary in flux between measurements can be susceptible to
enhanced proper motion errors.
We also find that for an on-source integration time of one hour, photon noise 
is larger than or of comparable magnitude to the confusion-induced position error for
bright targets and dominates over proper motion error in most cases.
\end{abstract}




\section{Introduction}

The Space Interferometry Mission (SIM) will allow astrometric measurements 
that are several hundred times more accurate than currently possible at 
optical wavelengths \citep[e.g.\ HIPPARCOS;][]{per97}. The key to this
giant leap in precision lies in exploiting the remarkable wavefront
stability in space via optical interferometry. SIM promises to achieve
$\mu$as accuracy for astrometry on objects as faint as $\rm m_v \lesssim 20$. 
At these faint levels, the presence of even fainter stars inside the
astrometric field of view is likely to perturb the astrometric measurements. 
Such ``confusion'' errors could limit the astrometric accuracy achievable with 
SIM.  
Some examples for SIM key projects\footnote{For a summary of the recently
selected SIM key projects, see 
${\rm http://sim.jpl.nasa.gov/ao\_support/ao\_abstracts.html}$}
which are likely to be affected by the presence of additional 
``background'' objects are 
(i) the determination of distances and ages of globular clusters, 
(ii) mass estimates for stellar, remnant, planetary, and dark-object masses
via astrometric micro-lensing, and
(iii) dynamical studies of external galaxies.

In this paper, we explore the effects of confusion on typical SIM 
astrometric measurements. More specifically, we model a number of 
target fields which we expect to be typical for the above projects.
These fields cover a fairly wide range in the degree of crowding, from the
densely populated M31 disk through the moderately crowded LMC to a sparsely
populated field in the bulge of our Galaxy. The results of this paper
thus should be useful for most SIM key projects to gauge the effect of 
crowding on the astrometric accuracy of the respective measurements.

The term ``confusion'' is most commonly used in connection with the
accurate measurement of target \textit{amplitude}, and has a rich history
in the context of the determination of the ``log(N) - log(S)'' source count
relation in radio astronomy and the application of radio source counts to
cosmology\footnote{\citet{mill84} has given an interesting retrospective of
the controversy between the Australian and Cambridge results on this
subject in the 1950's.}. Faint sources in the background which are within
the FOV of the telescope (i.e.  in the ``beam'') are a
source of noise which does not reduce with further integration.  Their
contribution to the estimates of source amplitude have by now been
thoroughly understood \citep{sch57,con74,fra82}. Confusion errors in
astrometry from imaging surveys have also been a subject of interest of
late \citep{hog00}.

In the case of astrometry with SIM, although the source of the noise is
still faint background sources within the FOV, we are especially concerned
with the effects of this confusion on the target \textit{position}, taken
in the context of an interferometric measurement of fringe phase.  We
investigate the role of instrumental parameters such as the FOV, the system
bandwidth, and the repeatability of the aperture re-pointing on the
confusion error.  The prominent role that interferometric techniques are
likely to play in future space-based astronomy serves as ample motivation
for this study.

The primary advantage of astrometry from space over ground-based
interferometric methods is that atmospheric turbulence is
avoided. Space-based interferometers are essentially phase-stable
instruments. Phase referencing within a fairly narrow field of view can
be used to overcome some of the atmosphere-induced errors from the
ground, but this is limited to narrow angle astrometry. 
Phase referencing using a star in the field of view also imposes
sensitivity limits on ground based interferometers. Low surface
brightness or extended objects are difficult targets for fringe phase
tracking from the ground. Space interferometry therefore offers a big
advantage for wide-angle astrometry and can reach fainter objects than
are possible from the ground. For example, the Navy Prototype Optical
Interferometer \citep[NPOI:][]{arm98} will achieve mas accuracy 
for wide-angle astrometry from the ground using $35\>{{\rm cm}}$ apertures and
baseline lengths of 19 to $38\>\rm m$.   SIM promises $\mu$as accuracy
with similar aperture sizes, and shorter baselines (up to $\sim 12\>\rm m$).
The limiting magnitude for NPOI is expected to be of the order of
${\rm m_v} = 7$ as compared to the SIM limiting magnitude of ${\rm m_v} \approx 20$. 

A tutorial covering the basics of astrometry with SIM can be found in
\citet{sha99}.  Briefly,
SIM uses interferometry to measure the angle between two stars.  The
position of a delay line which results in a ``white light'' or
zero-path-difference fringe is measured to a precision of $\approx0.1\>{{\rm nm}}$
(corresponding to a few $\mu$as in the sky) on a \textit{reference}
star.  The basic astrometric observable is the relative position of the
delay line with respect to this ``fiducial'' when a white-light fringe is
obtained on the \textit{target} star.  This measurement can directly be
translated to an angle in the sky between the two objects, projected
perpendicular to the baseline orientation.  The measurement is done for two
mutually perpendicular orientations of the SIM baseline, thereby yielding
the true angle between the reference and target sources.  The orientation
of the baseline in a reference frame is determined by observing selected
``grid'' stars, and ``guide'' stars are used to maintain the attitude of
the spacecraft to the high accuracy required.

Finding the position of the delay line for a white light fringe is the
basis of astrometry with SIM. This in turn involves measuring the phase
or position of the fringe pattern accurately. At least three sources of
error can occur: 1) the source structure itself can change   over the
course of a measurement or between measurements; 2) objects in the FOV
other than the target star can add to the fringe pattern, which leads to 
a modified fringe phase and thus to errors in the measured position of the star,
and 3) errors in the pointing of the siderostats can alter the FOV (for
repeated visits to the target) and hence the background sources
included in the FOV. This causes errors in the measured relative position of the
target between visits, which reflects as a proper motion error. In this
paper we address the latter two sources of error in detail. We also comment briefly on the first
source of error, namely, proper motion
error induced by the source varying in flux, a situation that might
be of concern e.g. in microlensing studies.

In \S 2 we present an analytical approach which provides a simple mathematical
description for confusion errors and a way of arriving at a first estimate for
the magnitude and scaling for such errors.
Section 3 describes the various steps involved in the numerical simulations 
we have carried out to accurately determine these errors. We 
present the results of these simulations for the specific fields we have
considered in \S 4 and \S 5. Section 6 is a brief summary.  
\section{Confusion errors in interferometric astrometry}

Confusion in the present context is the error introduced into the fringe phase of a
fairly strong signal (the target star) when superposed on a set of fringes from
weaker, randomly distributed sources (the background stars in the FOV).
When dealing with the fringe phase and amplitude (the
complex-valued fringe visibility), phasor notation is well suited to depict the situation
\citep{ryl59}. The errors in the measured fringe phase for a SIM
astrometric measurement in a crowded field can be derived from the well-defined 
statistical properties of the sum of a strong constant phasor plus
a weak random phasor sum \citep{goo85}.  A schematic representation is
given in Figure 1.  For a noiseless measurement, the fringe visibility of a source
at the phase center of the FOV is represented by a phasor of amplitude S
and phase zero\footnote{this is the reference phase and can be set to zero without
loss of generality}, i.e. lying on the real axis.  Background sources perturb
the fringe phase and are represented by the small amplitude, random phase
phasors at the tip of the strong phasor.  The loci (for different
realisations) of the tip of the phasor sum of the background are the circles which
form a noise ``cloud''.  The phase of the resultant R is now
$\phi$. The interferometric fringe phase $\phi$ for a given baseline B and wavelength $\lambda$ is given by 
\begin{equation}
\phi = \frac{2\pi \rm B\sin{\theta}}{\lambda} 
\end{equation}where $\theta$ is the angular distance of the
source from the line perpendicular to the baseline orientation.  
It follows that the root mean square value for $\phi$ represents the error in phase and hence
in the measured position ($\theta$) of the target.  In the case of background sources
which are weak compared to the target star, it can be seen from Figure 1
that $\phi$ is primarily decided by the imaginary part, Im(R), of the random phasor
sum.  On a subsequent visit to the field, a slightly different set of
background sources contribute to the random sum because of pointing errors,
and the total resultant is now denoted by the dotted line (R'), with a phase
$\phi'$.  The difference $\phi' - \phi$ represents the error in measured
proper motion between the two visits.  Here again it is the imaginary part
of the difference between the random sums, Im(R'$-$R), which dominates the error. 

The position and proper motion error caused by confusion
alone can be analytically estimated from the
statistics of the random phasor sum involved. We can estimate the
standard deviation of the phase ($\phi$ in Figure 1) of a fringe produced by the target
star from

\begin{equation}
\sigma(\phi) = \sqrt{\Sigma(\rm {X_i})^2/(2\rm S^2)}\>{\rm radians},
\end{equation}
where ${\rm X_i}$ are the intensities of the background
sources (the amplitudes of the weak random phasors), and S is the target
star intensity \citep[the strong constant phasor;][]{goo85}.  This quantity
is the confusion error in the position of a source in a crowded field.  The
error in proper motion ($\phi' - \phi$ in Figure 1) due to
pointing inaccuracies is given by the ratio ${\rm Im(R'-R)}/\rm S$ and scales 
inversely as the target intensity.
Equation (2) shows that the position error $\sigma(\phi)$ also
scales inversely with the target intensity S.   

The above description provides a simplified picture for the nature
of confusion in interferometric astrometry. In practice, the situation is more
complicated because of photon noise contributions and effects of the finite
bandwidth. The finite bandwidth efffectively reduces the FOV. Sources at larger
angular distances from the line perpendicular to the baseline orientation suffer larger decoherence.
To clarify, recall that equation (1) gives the interferometric phase for a single
wavelength $\lambda$. For a finite band, the amplitude of each of the random
phasors in Figure 1 results from vector-adding all the phasors corresponding
to each wavelength in the band. This results in a reduction of the amplitude, an
effect which is larger for phasors with larger phase angles, i.e. sources at larger distances
from zero delay. In essence, this is the ``delay beam''  which 
progressively decreases the contribution of background sources to the resultant
fringe as they are moved further away from the line perpendicular to the baseline orientation.  
In order to account for these effects, we have simulated typical
SIM fields and numerically estimated the total astrometry error from confusion and 
photon noise with bandwidth decorrelation taken into consideration. In the absence of
detailed information on detector characteristics, we have not modeled read noise
or other similar sources of error. The rest of this paper 
describes these simulations and their results.
In an earlier study, we have
estimated both position and proper motion errors using the statistical
results described above for the specific example of M31 \citep*{raj99}.  The 
earlier results are consistent with
those achieved with the full simulation described here. 

\section{The simulations}
A pre-requisite for estimating errors from confusion effects is to build a
model of the sky as seen at the SIM resolution. 
Our model of the background seen in a typical SIM FOV
is based on HST archival images taken through the Wide~Field~Planetary~Camera~2
(WFPC2). The resolution afforded by the HST images is important to accurately
model the changes in the background structure for SIM pointing errors which are
of the order of a few tens of milliarcseconds. In this section we detail the steps involved in the
simulation. We account for the effects of the point spread function of the 
siderostats, size of the field stop and the bandwidth used. 
We chose the fields to be typical examples for the kind of
targets that SIM is likely to study extensively, and we discuss each
example in detail in \S 4 and \S 5.  

The simulation involves the following basic
procedure.

\begin{itemize}
\item[1.] In approximating the SIM FOV, we assume that the astrometry measurements are carried
out using a baseline of $10\>\rm m$. Using the HST image to
constrain the total flux in the SIM FOV, a model sky field at the resolution
afforded by this baseline ($\sim 10\>{{\rm mas}}$ at a wavelength of
$600\>{{\rm nm}}$) is constructed. An area large enough to accomodate
the effective SIM FOV is chosen at random on the HST image.
The simulated FOV is constructed with 5 mas pixels (assuming
Nyquist sampling for SIM) and
the total flux is redistributed in this field among stars randomly drawn from a 
luminosity function (LF).  For all our sample fields, we have used
the LF for the solar neighbourhood from \citet*{yos87} with
appropriate limits which are specified in the discussion for each field. 
Each star is then put down as a $\delta$~function at random positions within that FOV. The SIM astrometry measurement
is simulated for a number of such locations.

\item[2.] Having modeled the background as seen at SIM resolution,
the next step involves calculating how each of these sources
contribute to the measured fringe amplitude and phase. There
are three parameters which define the relative strength of their
contributions: the PSF of the individual SIM siderostats,
the size of the field stop in the optical path, and the
bandwidth decorrelation effect. While the first two affect
the total number of photons available from a source depending
on its position in the FOV, the third modulates the coherence
of the light from a source, thereby affecting its contribution
to the fringe pattern without changing its photon flux. 
To account for the siderostat PSF and the field stop,
we multiply the FOV by a function obtained from the convolution
of the Airy disk (at the center wavelength) of the siderostats with the 
field stop function. This
function is centered on the FOV. The actual area of the image
considered is greater than the FOV to account for contributions
from sources beyond the edges of the field which ``leak'' in
because of diffraction effects. The PSF is calculated for a siderostat
diameter of $30\>{{\rm cm}}$ (the currently expected size for the SIM apertures). The field stop shape
has as yet not been decided, and we have assumed a square stop.

\item[3.]We discuss the effects of bandwidth decorrelation (the ``delay beam'') for bands of
$12\>{{\rm nm}}$ and $400\>{{\rm nm}}$ centered at the WFPC2 filter center.
The expected total bandpass for SIM is $\sim 400\>{{\rm nm}}$ (from
$500\>{{\rm nm}}$ to $900\>{{\rm nm}}$) with the fringes dispersed into individual channels having a
width of $\sim 6\>{{\rm nm}}$. Hence  the lower
value we have used is appropriate for summing two channels
whereas the higher one is applicable when most of the channels are
co-added. Binning channels to estimate the fringe phase is
done post-observations and these two cases correspond to the two
limits for the number of channels addded together.
The location of the center of the delay beam on the sky is decided  
by the positioning of the delay line. The
nominal value for the error in delay line position is $\sim 10\>{{\rm nm}}$
which corresponds to roughly 2$\%$ of the fringe width at the
shortest wavelength. We neglect this source of error
in the simulations, i.e. the position of the delay beam is considered stable 
and only the FOV is ``jittered'' to account for pointing errors.
The delay beam corresponds to a 
strip in the sky perpendicular to the given baseline
orientation. Its profile is given by
the Fourier transform of the bandpass function. Here we assume a rectangular bandpass.
The decorrelation is implemented by simply vector averaging the
visibility values at the (u,v) coordinates corresponding to 
the $10\>\rm m$ baseline at the center wavelength of each channel in the
band considered. The effect of the finite channel width is
ignored. 

\item[4.]Finally, we account for the photon noise contribution to the errors.
This is done by simulating the fringe visibility measurement 
as described in \citet{bok99}. In brief, the delay line is stepped
through four different values of delay around the white light or zero
path length difference setting. The  photon counts in each of the 
four delay bins for each channel are then used to calculate the fringe
visibility. This procedure is simulated for typical bandwidths and
integration times and the counts in each bin 
modified according to Poisson statistics
to account for the photon noise. To obtain estimates
for confusion noise alone, this feature can be switched off.  

\item[5.] The preceding steps allow us to ``measure'' the
fringe phase corrupted both by confusion
from other objects in the FOV and photon noise. 
To arrive at the fringe phase, we Fourier transform the simulated
FOV and pick out the phase of the Fourier component 
corresponding to the  baseline used ($10\>\rm m$).  

\item[6.]To calculate the rms error in the measured position of a target
star, we determine the fringe phase of the target star
which is introduced into the FOV at the phase center (and
therefore expected to have zero phase).  
By simulating the measurement at a number of different
locations (100 in this case) on the image and measuring the spread of values for
the measured phase, we can estimate this 
error. In the absence of photon noise, this corresponds to the position error
for the target because of confusion from background objects.

\item[7.]For a proper motion measurement, the same field would be visited  more
than once and changes in the position of the
target are measured. For each visit, the FOV will be slightly different
because of pointing errors. The relative change in the phase of
the measured fringe (because of the different distribution of background sources)
manifests as an error
in the measured proper motion. To simulate this error, steps
1 through 4 are done, and the {\em change} in phase ($\Delta \phi$)
of the $10\>\rm m$ Fourier component as the FOV is shifted  
by a small amount is measured. This is repeated at each location
for a range of values for the pointing error and the spread
in the values of $\Delta\phi$ is computed as the estimate for
error in proper motion because of pointing errors 
in a crowded field. The shifts are carried out in the direction
perpendicular to the delay beam orientation. This is the direction
in which the maximum change in background is expected and 
provide a worst case estimate of confusion-induced proper motion errors.

\end{itemize}

\subsection {Important instrument parameters}
The important instrument parameters which we assume for the simulations are:
\begin {itemize}
\item[] Baseline length: We have used a baseline length of ${\rm 10\>m}$ for all
results quoted here.
\item[] Mirror size: The siderostat diameter assumed is $30\>{{\rm cm}}$.
\item[] Throughput: We have assumed an overall throughput of 0.3.
\item[] Field stop size: The field stop diameter (or side for a square stop) is expected to be in the
range of 0.3\arcsec~(the FWHM of the PSF at $600 \>{{\rm nm}}$) to 1.0\arcsec.
\item[] Bandwidth: SIM is expected to have a wavelength range
of 400 to $900 \>{{\rm nm}}$, with a resolution of $\sim 6 \>{{\rm nm}}$. We specify
the bandwidth used for each example presented here.
\item[] Pointing accuracy: On a bright source, where the
star trackers can be used to guide the instrument, the pointing
accuracy is expected to be of the order of $10\>{{\rm mas}}$. This is
the accuracy to which a given pointing can be
repeated. The jitter on any individual pointing will be less
than this value. For a faint source, the instrument will make use of
information from the guide interferometers to maintain
pointing. In this case the accuracy for revisiting the field
is expected to be $\sim 30\>{{\rm mas}}$. We show results for 
a range of pointing offsets. 
\end{itemize}
 
In the case of the M31 field, we performed the simulations using a range of likely values for some of
the crucial design parameters for SIM which are either not yet fully
specified or are likely to vary depending on the observation strategy
chosen.
These include the bandwidth,
the field stop size, and the pointing accuracy, and we have attempted to
establish the trends in astrometry error for different values
of these. In the case of the LMC and Galactic bulge fields, we show results
for a given set of the most probable values for these parameters.

\section{Results for M31}
We used archival HST images (Program ID 5971, PI: Griffiths, R.E) of a region of the M31 disk taken with the
Wide Field Planetary Camera (WFPC2) through the F606W filter
(mean wavelength $\sim 584 \>{\rm nm}$) to model the background seen
by SIM. We have chosen this field (Figures 2 and 3) to be located in the disk
of the galaxy in an area where a typical astrometry measurement
for measuring ``rotational parallax'' might be carried out.
Measuring the distance to M31 using this method
is an idea which dates back to the beginning of the century.
It involves obtaining the proper motion of individual
stars in the disk at several locations. When combined with radial
velocity measurements, this yields an independent estimate of the
distance to the galaxy, as well as its rotation curve and inclination
to the line of sight.  For the first time, SIM will provide the necessary
precision to achieve these remarkable goals. The required accuracy of
proper motion is of the order of 5 $\mu$as per year  for
stars of magnitude ${\rm m_v} = 16$ in the M31 disk \citep{oll00} which is within
reach of the sensitivity limits specified for the mission \citep*{all97}.\footnote{For 
SIM science requirements see also the report of the SIM Science Working Group
at http://sim.jpl.nasa.gov/library/technical\_papers.html}
However, in the case of crowded fields, the effect of confusion 
needs to be studied in detail and will have an important role in
defining the actual limits of astrometric precision. 

\subsection{Source Model}
The image is from coadded multiple exposures and has a net
exposure time of $5600\>\rm s$. It has been subjected to the
pipeline calibration procedure. The multiple exposures have been
used to remove cosmic ray events.  The resolution is $\sim 0.1\arcsec$
per pixel. Each of the three WFPC2 chips covers $\sim 80\arcsec$
of the sky. 

To model the FOV at SIM resolution (step 1 of previous section) we
use the LF for the solar neighbourhood from \citet{yos87}.
This LF has been extrapolated at the high luminosity end to an absolute 
magnitude (V band)
of $-$8.5.
The linear extrapolation uses a logarithmic
slope of $-$5.4 which is consistent with the measured slope of
the LF in the relevant magnitude range in M31 \citep{hod88}.
The high end limit corresponds to an apparent magnitude of 16
at the distance to M31 \citep[distance modulus of 24.5;][]{sta98} which is
the proposed typical magnitude of a target star for the rotational
parallax observations. 
For the low luminosity limit, we use an absolute
V magnitude of 6.0 (${\rm m_v} = 30.5$), which is well below the expected 
SIM sensitivity limit of ${\rm m_v} = 20.0$ \citep{all97}. The integrated flux
flattens out at this magnitude indicating that the flux contribution
from fainter stars is negligible.

For this field as well as all other simulations described in this paper, we have investigated the
error both in position and proper motion.
Two important parameters that influence the
magnitude of these errors are the extent of the FOV and the effective bandwidth
used. We have varied these parameters to gauge their influence on
the accuracy of proper motion measurements with SIM. This will
be relevant for both the SIM design and observing strategies adopted.

\subsection{Position error}
We briefly describe here
the results of our simulation on the M31 field to gauge position error (see Table~\ref{tbl-1}). 
Here and in the following sections, we quote errors for the assumed baseline of $10\>\rm m$ and
wavelength of $600\>{\rm nm}$. 
The target star in our simulation is
at the phase center and in the absence of noise should have a measured fringe phase
of zero. For a bandwidth of $400 \>{\rm nm}$ (from $500$ to $900 \>{\rm nm}$) and a field stop size
of 1\arcsec, the rms deviation from this value 
because of background sources alone (no photon noise)
is of the order of 0.7 $\mu$as for a target
star of ${\rm m_v = 16 }$ and is a measure
of the position error.
A narrowband ($12\>{\rm nm}$) case shows a significant increase in the
position error since the bandwidth decorrelation is now much reduced (the ``delay beam'' is
now much broader) and a larger number of confusing
sources contribute to the error phasor. 
The number of contributing sources is now limited by the
size of the field stop. We find that the error is reduced by a 
factor of 2 when the
field stop size is decreased to 0.3\arcsec. In some trials, the (weighted) FOV includes sources which are of
comparable magnitude to the target and lead to comparatively large errors. 
Identifying these outliers through fringe fitting (\citet{dal01} discuss fringe 
fitting for some specific examples of LMC fields) or avoiding such fields through
pre-selection can cause the error to be reduced by a factor of 3 or more as
discussed in more detail in the following section. 
The errors scale approximately as the inverse of the flux ratio (see Figure 7), in
agreement with the analytic expression for confusion error in \S 2.
It is evident from Table \ref{tbl-1} that position errors are quite significant when
compared to the required SIM accuracies for all except the very brightest targets.
\subsection{Proper motion error}
 
The results for the proper motion error are also summarised in Table~\ref{tbl-1}.
Figure 4 shows the proper motion
error for a range of
pointing offsets for a 16th magnitude target star. These are estimated as described in \S 2 (step 7)
by taking the standard deviation of the change in phase for a given
pointing offset  over a number of randomly picked locations on the HST image.
The simulations do not include photon noise and the error is from confusing
sources alone.     
Figure 4 is for a bandwidth of $400\>{\rm nm}$ (from $500-900\>{\rm nm}$) which covers
most of the SIM spectral range and the field stop is a square of side 1\arcsec. 
Here, the large bandwidth decorrelation effect dominates in deciding the effective
FOV and the confusion.
The value for proper motion error at  0.01\arcsec pointing offset is $\sim$ 0.006
$\mu$as. This is negligible compared to 
the specified instrument accuracy of $\sim$ 1 $\mu$as and the required
proper motion accuracy of $\sim$ 5 $\mu$as for the rotational parallax experiment \citep{oll00}. 
Figure 5 (triangles) is
for a bandwidth of $12\>{\rm nm}$ ($694-712\>{\rm nm}$).
The proper motion errors have increased by a factor of $\sim$ 10,
still small compared to the sensitivity required. The  error increases since
the bandwidth decorrelation is reduced and more sources 
contribute to the confusion. Since the delay beam does not dominate (unlike the broadband case) 
in limiting the extent of the effective FOV, the weighting function
derived from the Airy disk and field stop size (step 5 in \S~2) 
may play a role in the magnitude of confusion error. In Figure 5 we also
show the results when the field stop size is decreased to 0.3\arcsec.
We do not find a significant change in the proper motion error.
This is different from the position error which did decrease with
stop size.
The field stop size therefore can be chosen to maximize the throughput because confusion is unlikely 
to play a role in this design criterion.

The large position errors in Table 1 do not result in large \textit{proper
motion} errors since most of the position error is common between visits and
therefore cancels out. However, if the target flux were to change between visits, the
position errors (which scale inversely as the target flux) would not cancel out and
would indeed cause proper motion errors. A typical example 
would be a proper motion measurement for a microlensing event, where the brightness of the
lensed object can easily change by a magnitude or more. For a 19th magnitude target 
in the LMC, the position error is $\approx$ 3 $\mu$as (\S 5). If the target
dims to 20th magnitude on a subsequent visit, the position error scales to $\approx$ 7.5 $\mu$as.
The difference or ``proper motion'' measured would be therefore 4.5 $\mu$as from this
effect alone. In comparison, the typical signal in such experiments would only be a few $\mu$as.
We note the serious nature of this problem here and hope to address possible remedies in
a future publication.

\begin{deluxetable}{cccccc}
\tablecaption{Summary of results for M31 \label{tbl-1}}
\tablewidth{0pt}
\tablehead{
\colhead{Bandwidth} & \colhead{Field stop}   & \multicolumn{2}{c}{Position Error } & 
\multicolumn{2}{c}{Proper Motion Error } \\
\colhead{[nm]} & \colhead{[as]} &\multicolumn{2}{c}{[$\mu$as]} &\multicolumn{2}{c}{[$\mu$as]} \\
\colhead{}&\colhead{}&\colhead {${\rm m_v = 16}$}&\colhead {${\rm m_v = 19}$}&
\colhead {${\rm m_v = 16}$}&\colhead {${\rm m_v = 19}$}
}
\startdata
400 &1.0 &0.7 & 9.9 &0.01 &0.1 \\
12 &1.0 & 3.2 & 53.1 & 0.06 & 0.9\\
12 &0.3 &1.9 & 30.0 & 0.05 & 0.8\\ 
\enddata



\tablecomments{ The proper motion errors are for a pointing offset of 0.01\arcsec, the value 
adopted as the nominal pointing error for the SIM siderostats. The phase errors in
radians have been converted to angle in the sky using a baseline of $10\>\rm m$, for a
central wavelength of $600\>{\rm nm}$. No photon noise is
included. Note that the errors scale inversely with target flux. 
}

\end{deluxetable}

Figure 6 shows a histogram of the error in proper motion 
measured with a $10\>\rm m$ baseline for a 16th magnitude target star as
the FOV is offset by 0.01\arcsec  (the pointing error) at 100 random locations
in the HST image. The histogram has a central peak and some
outliers. These large deviations in phase are a result of background
stars of comparable magnitude to the target in some of the
locations and these contribute significantly to the standard deviation
of the samples and hence to the error estimates. It is likely
that careful selection of SIM target fields will avoid this problem. We
show in Figure 6 a Gaussian fit to the central peak of the distribution to 
estimate the proper motion error (the sigma of the Gaussian). All values for 
both position and proper motion errors listed in this paper are derived in this fashion. 
The directly computed standard deviation is a factor 3 to 5 greater than these values because
of the outliers. 
However, the scaling of the errors with 
target star magnitudes or pointing errors are not affected.
\subsection{Scaling with target flux}
From equation (2) both the position and proper motion errors
are expected to scale inversely with target brightness.
In Table \ref {tbl-1} we show the error values for targets
of ${\rm m_v} = 16$ and 19 to demonstrate this. In addition, Figure 7 shows the scaling 
of both astrometry and proper motion errors with the inverse flux ratio, normalised to the value for a
target star of ${\rm m_v = 16}$ and a bandwidth of $400\>{\rm nm}$.
The scaling is approximately linear until the target flux drops by
a factor of $\sim$ 15 (${\rm m_v=19}$). The simulations show that as the target flux drops beyond 
this value the confusion error starts
to flatten out as the ``noise cloud'' (Figure 1) starts to be of comparable
amplitude to the target phasor itself. In this limit the linear scaling law is no longer applicable. 
Figure 8 shows the scaling of the proper motion error with magnitude for the $12\>{\rm nm}$ bandwidth
case. The scaling is shown for field stop sizes of 1.0\arcsec and 0.3\arcsec. Here also the scaling is
linear for moderate flux ratios. 
\subsection{Photon noise}
We show in Figure 9 the proper motion error histogram from a simulation
with photon noise contribution included. The input parameters are the same
as in Figure 6 (which does not have photon noise), with an integration time of one hour.
The photon noise clearly dominates the error in measured phase and is
a few hundred times larger than the confusion-induced proper motion error. For the
case shown, the measured phase distribution has a sigma of  $\sim$ 0.001
rad, which implies an error in the measured proper motion of $\sim$ 2.0 $\mu$as (as in all cases, this
is for a baseline of $10\>\rm m$ and central wavelength $600\>{\rm nm}$). 
The size of the field stop plays a role in the photon noise
contribution.  A stop of diameter 1\arcsec allows 80 to 90$\%$ of the target
flux through for the $500$ to $900\>{\rm nm}$ wavelength range and a $30\>{\rm cm}$ aperture. Further increase
of the stop size will increase the background flux level without
adding significantly to the source flux and the fringe visibility
is reduced. This directly increases the phase error as
$$\phi_{rms} = \frac{\rm C}{2\pi\rm V\sqrt {\rm N}}$$
where C is a constant, N the photon number and V the fringe visibility \citep{sha77}.
However for the cases we have studied the target
flux is high enough that the background 
only contributes a small fraction of the overall flux 
and hence
this effect is well below the statistical errors in our simulations.   

The magnitude of the error in the measured absolute position of the target
from photon noise is similar to that in the above case. 
Hence it is clear that the confusion-induced position error (ref Table \ref{tbl-1}) is of comparable 
magnitude to the photon noise error for the brightest targets in M31. Since confusion error 
scales inversely as the flux
and the photon noise scales inversely as the square root of the flux, the position
error can be dominated by confusion for weaker targets.  

\subsection{Some approximations}
Before presenting the results for the other fields, we discuss 
some important approximations assumed in the simulations and possible
consequences.  
 
In principle, the estimates of noise from confusing sources could be affected by the photon
noise in the HST image we use to model a SIM FOV. Since we attribute all
the flux to point sources, photon noise induced variations can introduce
false structure into the simulation. However, the typical pixel-to-pixel variance measured
in the deep exposure images of the M31 disk that we use are 8 to 9 times larger than 
expected from Poisson fluctuations of the photon counts and so the structure in the image is
real. The formal errors (1 $\sigma$) on the gaussian fits (with statistical weighting for 
the bin counts) to the phase deviations are $\sim$ 10$\%$.
The formal errors are consistent with the variations between different runs of the simulation. 
We therefore take this value as
the error in our estimates for confusion noise. We note that eliminating fields
with bright background sources can systematically reduce the error. 

We model the position offset (caused by pointing errors) to be perpendicular to
the orientation of the delay beam on the sky, in order to fully gauge the
effect of the bandwidth on the confusion error. For the wide band ($400 \>{\rm nm}$) case,
this is likely to give larger estimates for the confusion than in the actual
case of pointing offsets in random directions. For the narrow band, the
bandwidth decorrelation has minimal effect irrespective of the adopted
direction for pointing offsets.  

The only instrumental errors we have considered for this analysis are 
pointing imperfections. In the absence of detailed information of other sytematic errors, 
we currently believe that pointing is indeed the major factor for confusion issues. 

\section{Results for the LMC and the Galactic Bulge}
The magnitude of confusion errors in SIM astrometric measurements is clearly affected by many factors. 
For example, the distance to the
target field and the structure of the background play important roles. For this reason, it is 
difficult to extend the M31 results to other fields in any simple fashion. To gauge
the effect of distance to the target field on the confusion problem, we carried out simulations
on typical fields in the LMC and our own galaxy. Together, these three
cases form a fairly representative sample of distance, surface brightness and the degree of crowding 
for most targets that SIM is likely to observe. Hence these results are useful in estimating 
the extent of confusion for most SIM programs. The main results are described below 
for comparison with the M31 results.
  
Fields in both the LMC and the Galaxy will be observed by SIM for a variety of reasons. For example, the
fields we have chosen are typical candidates for  microlensing events.
The lensing event  causes a shift in position of the centroid of the
lensed star. This shift along with the measured light curve can be used to
obtain distances to the lensing and lensed object as well as the mass 
of the lens. These experiments try to detect a relative change
in position of the target star over repeated visits to the field. The
presence of other sources in the FOV, coupled with pointing inaccuracies
cause proper motion error. For both the
LMC and Bulge microlensing events, the predicted centroid shift is of the
order of a few $\mu$as. In both cases, we have adopted a V magnitude
of 19 for the target star. Since this is a fairly weak target (the
SIM sensitivity limit is $\sim$ ${\rm m_v = 20}$), the experiment will
aim to maximise the signal from the target. We therefore
use a wide band ($400\>{\rm nm}$) and a field stop size of
1\arcsec for these simulations. 
Here we aim to estimate the confusion noise alone and do not include the photon noise contribution.
We have already demonstrated that even in one of the more
extreme cases of crowding (the M31 disk), 
photon noise is larger than the confusion-induced position error and dominates over the proper motion error for 
a one hour integration.
Hence the photon noise contribution to astrometry errors in the LMC and
the Galaxy (or indeed any other field) can be scaled
from that result using the total flux and standard Poisson statistics, neglecting
the contribution from confusion. 

Figure 10 shows the proper motion error histogram for the case of the LMC field and a pointing
error of 0.01\arcsec. The HST image (Program ID 5901, PI: Cook, K.) used here is
from  the Planetary Camera (PC), with the F814W filter (I band) and $500\>\rm s$
exposure time. We use the same LF to populate the simulated SIM
FOV as in the previous case, with upper and lower limits of ${\rm M_v} = -1.0$ and 15 
respectively. The histogram clearly shows that the proper motion error
has large excursions from zero for a few trials. As in the M31 case, this is
due to background stars of comparable magnitude to the target inside the FOV.
Since the target magnitude of ${\rm m_v = 19}$ \citep[${\rm M_v} = 0.5$ for a distance modulus of
18.5 for the LMC;][]{mad98} is much weaker than for the M31 case, the relative
strength of the background sources within the FOV is higher, resulting in large
deviations of the measured phase. Such fields should  be easy to identify and
this effect can be accounted for (for example with fringe fitting methods) to reduce the
error. If we neglect the outliers in the histogram, the proper motion error
is  negligible (0.02 to 0.03 $\mu$as). 
The position error for this case is $\sim$ 3 $\mu$as.
As described in \S 4, this could be of significance in the proper motion
measurement if the target were to vary in flux. 

For the Galactic Bulge, we have used an HST PC image (Program ID 7437, PI: Bennet, D.), taken with
the F555W filter (V band) and a short exposure time of $40\>\rm s$. In
this case, photon noise in the HST image could affect the simulated
background. To minimise this, we have only used those HST pixels 
with counts more than twice the rms noise level in the image.
The LF used has limiting absolute magnitudes of 3.5 and 15.5.
The target star has an apparent magnitude of 19 \citep[the distance modulus
used here is 14.6;][]{alv99}. Other parameters are the same
as in the LMC case.  
Figure 11 shows the proper motion error h istogram. 
Here also, except for instances of clear outliers, the proper motion error is
negligible and of the order of 0.002 $\mu$as. The position
error in this case is $\sim$ 0.2 $\mu$as.
\section{Summary}
We have estimated the error in measured position and proper motion arising from  sources
other than the target in the SIM FOV and the dependence of this error
on parameters like pointing accuracy and target flux. We have also discussed the effect of the bandwidth
and field stop size on this error. 

The error in position of the target star because of background 
sources is of the order of 1 $\mu$as for the M31 disk for the brightest
possible targets. 
The error increases linearly with decreasing target flux. As is evident from
Table \ref{tbl-1}, the position error is considerable for the narrow
band case and is quite significant for weaker targets in M31. The same
holds for the LMC and Galactic bulge as well.

Errors in proper motion are much smaller than the absolute position
errors. This is because proper motion is essentially a relative measurement and most
of the confusion induced position offset tends to cancel out between successive visits to the target.
For the case of a target star of apparent
magnitude ${\rm m_v} = 16$ in
the M31 disk, the proper motion error is a small fraction of the required sensitivity
for reasonable values of the relevant
parameters. 
The target here is among the
intrinsically brightest (M$_v = -8.5$) known stars. We show that the confusion-induced errors scale 
inversely as target flux and the proper motion error
can be significant for weaker targets in M31. For the LMC and Galactic bulge,
the confusion induced proper motion error is not a significant source of error even for targets close to
the sensitivity limits specified for SIM (when using most of the available
bandwidth). However, the fringe phase can be
corrupted by the occasional strong source within the FOV.
For the case of variable targets, the 
confusion induced position offsets will not be the same between visits and hence
will not cancel out in a proper motion measurement. This could lead to large proper
motion errors, comparable to the position errors themselves. 

Both for position and proper motion errors, significant reductions
can be obtained by: selecting fields without other
objects of comparable brightness; by identifying and
removing the contribution to the fringe phase from such objects; and
using all the available bandwidth for estimating the fringe phase.

The field stop size plays only a minor role for position errors in
the wideband case, since the narrow delay beam essentially limits the
FOV. For the narrowband case, there is a decrease in confusion-induced
position errors with decreasing field stop size. However, these errors
are still within tolerable limits for bright sources for a field stop
as large as 1\arcsec. A field stop of this size also maximizes throughput from
the source without including too many background photons (in any case, for our sample fields 
the photon noise contribution from
the background is small.) We find that confusion-induced proper motion errors are
largely insensitive to field stop size. 
   
The contribution from confusion to proper motion errors is in most cases much
less than that from photon noise for an integration time of one hour. The position errors 
from confusion however are of
comparable magnitude to the photon noise and can even dominate for some cases.  
\acknowledgements
We thank Ken Freeman, Dean Peterson, Steve Unwin, Roeland van der Marel and Neal Dalal for discussions
and suggestions. Rosa Gonzalez and Kailash Sahu helped us by providing access
to and information about the HST archival
images used in the simulations.    
The work described here was carried out at STScI with financial
support from the SIM project at the Jet Propulsion Laboratory (JPL).
\clearpage
\begin{figure}
\plotone{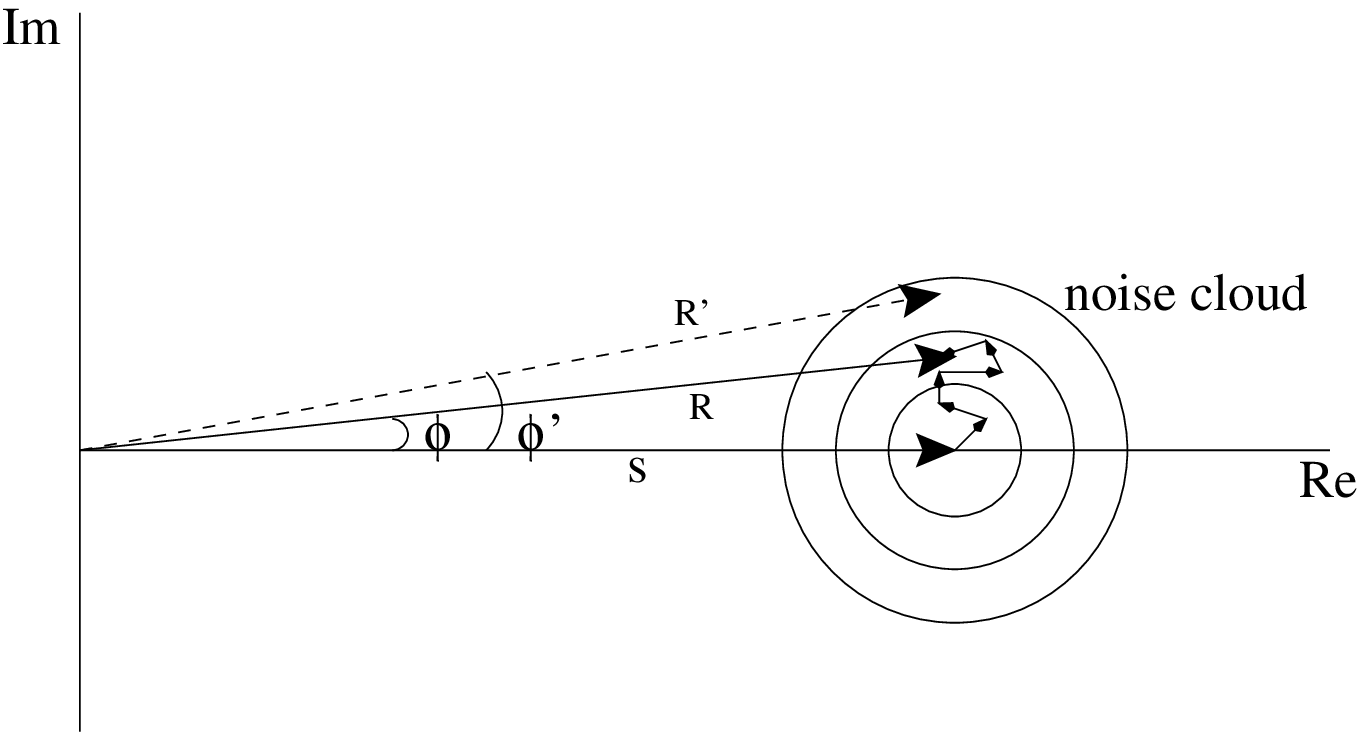}
\figcaption[f1.eps]{Schematic clarifying the phasor notation for
our confusion model. The target phasor S lies along the
real axis (reference phase set to zero). 
The random sum of the confusing
sources (denoted by the small amplitude phasors at the tip of
S) perturbs the phase measurement for S. The resultant R has phase 
$\phi$ (the position error). On a subsequent visit to the same field, pointing
errors can result in a slightly different phase ($\phi'$) for
the resultant (R'), causing a proper motion error. \label{fig-1}}
\end{figure}
\clearpage
\begin{figure}
\plotone{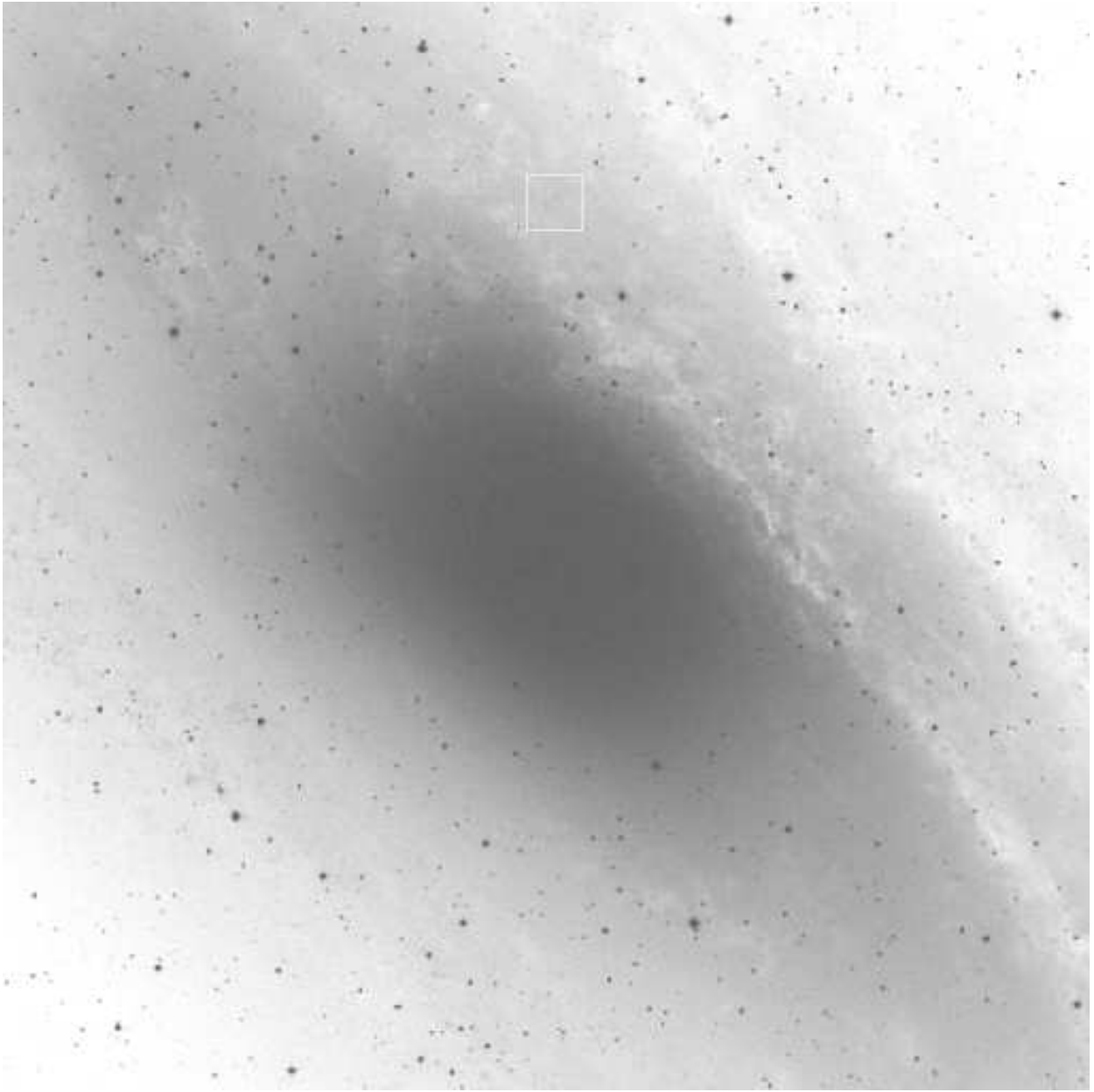}
\figcaption[f2.ps]{DSS image of M31 (30' $\times$ 30'). A typical
location chosen for our simulations is marked. \label{fig-2}}
\end{figure}
\clearpage
\begin{figure}
\plotone{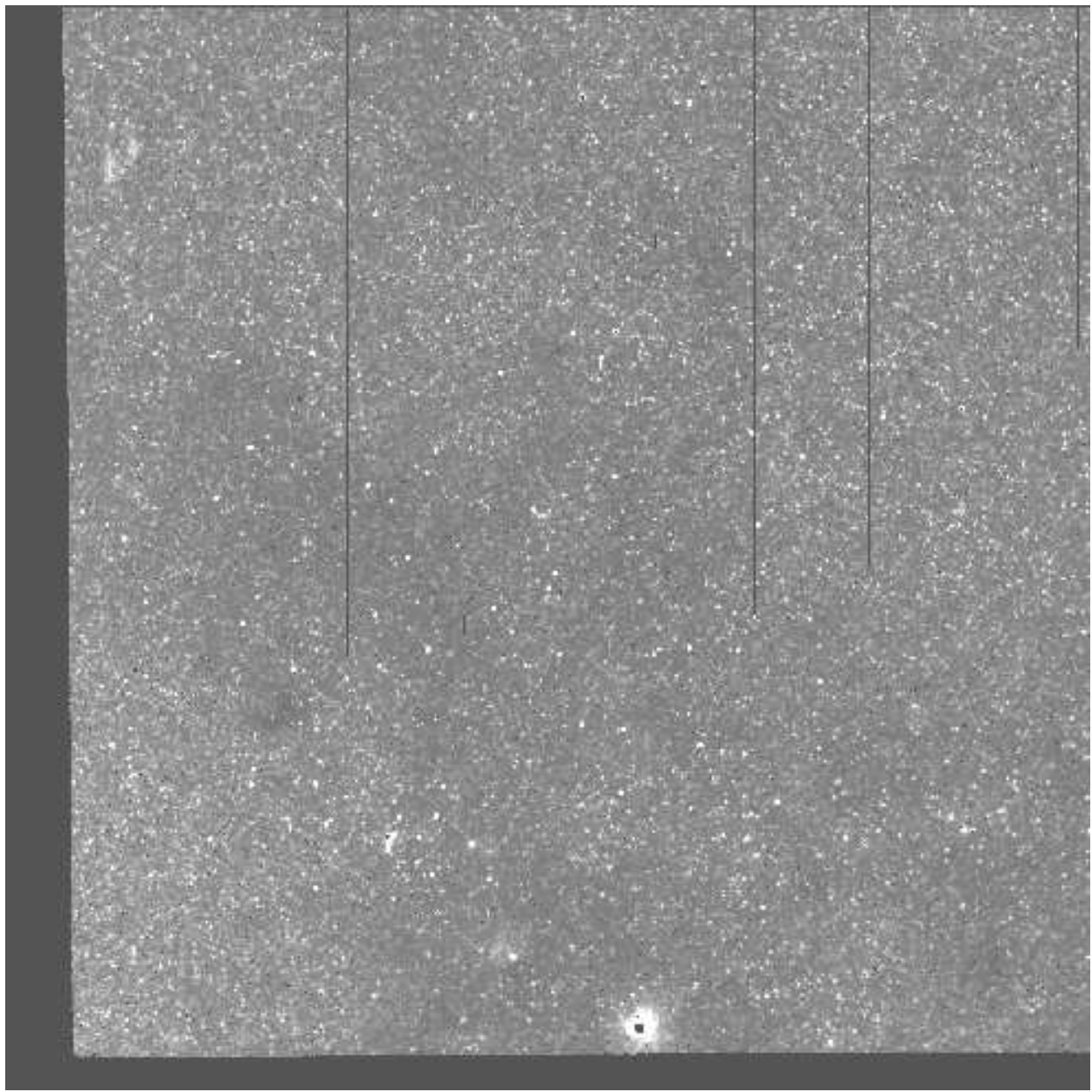}
\figcaption[f3.ps]{WFPC2 image of the location in the M31 disk shown in
Figure 2, approximately 1' to the north of the nucleus (F606W filter, 80\arcsec$\times$80\arcsec
with 0.1\arcsec per pixel). Several stars of ${\rm m_v=16}$ have been introduced to help
gauge the relative background surface brightness. \label{fig-3}}
\end{figure}
\clearpage
\begin{figure}
\plotone{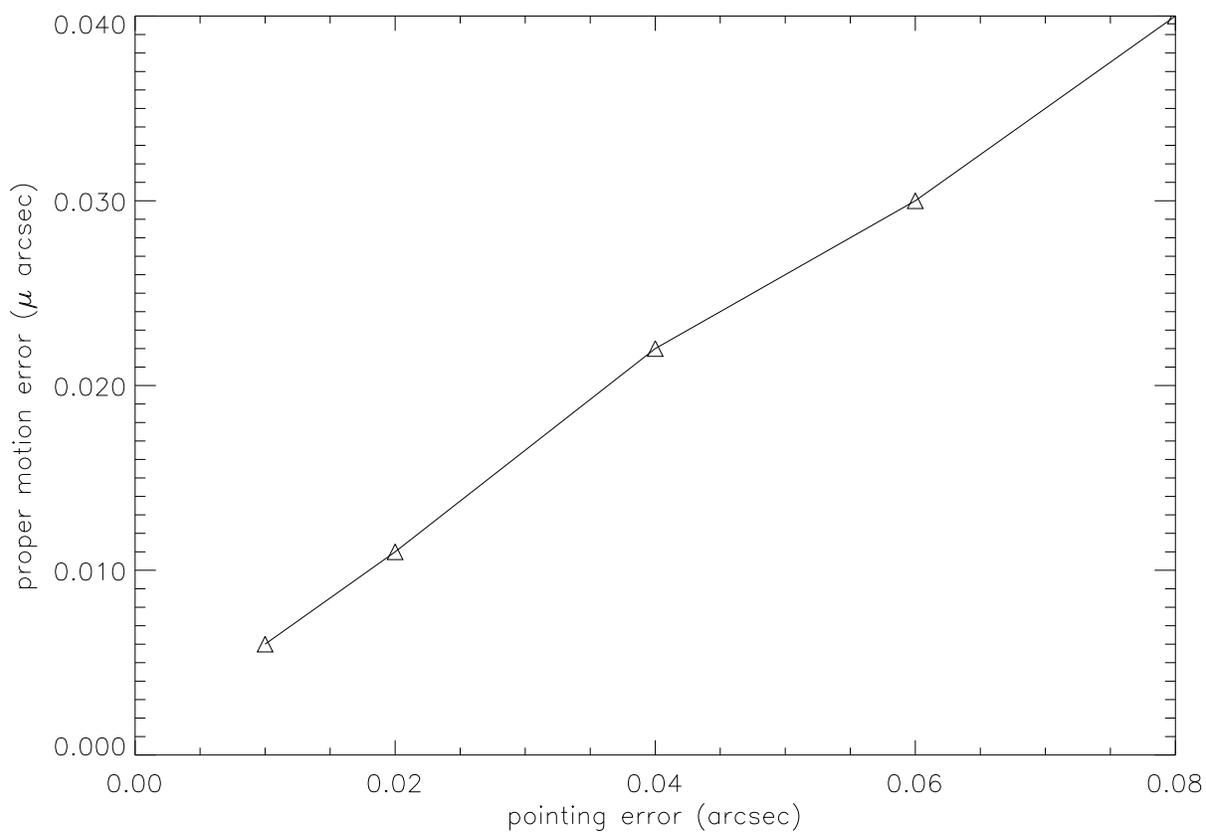}
\figcaption[f4.ps]{The proper motion error due to confusion, plotted against the
pointing offset for the M31 field, for an ${\rm m_v} =16$ target star. The bandwidth used is $400 \>{\rm nm}$ and
field stop size is 1\arcsec. The pointing accuracy for SIM is expected to be $\sim$ 0.01\arcsec. 
The proper motion error for other values of target brightness can be obtained
by scaling inversely as the target flux. \label{fig-4}}
\end{figure}
\clearpage
\begin{figure}
\plotone{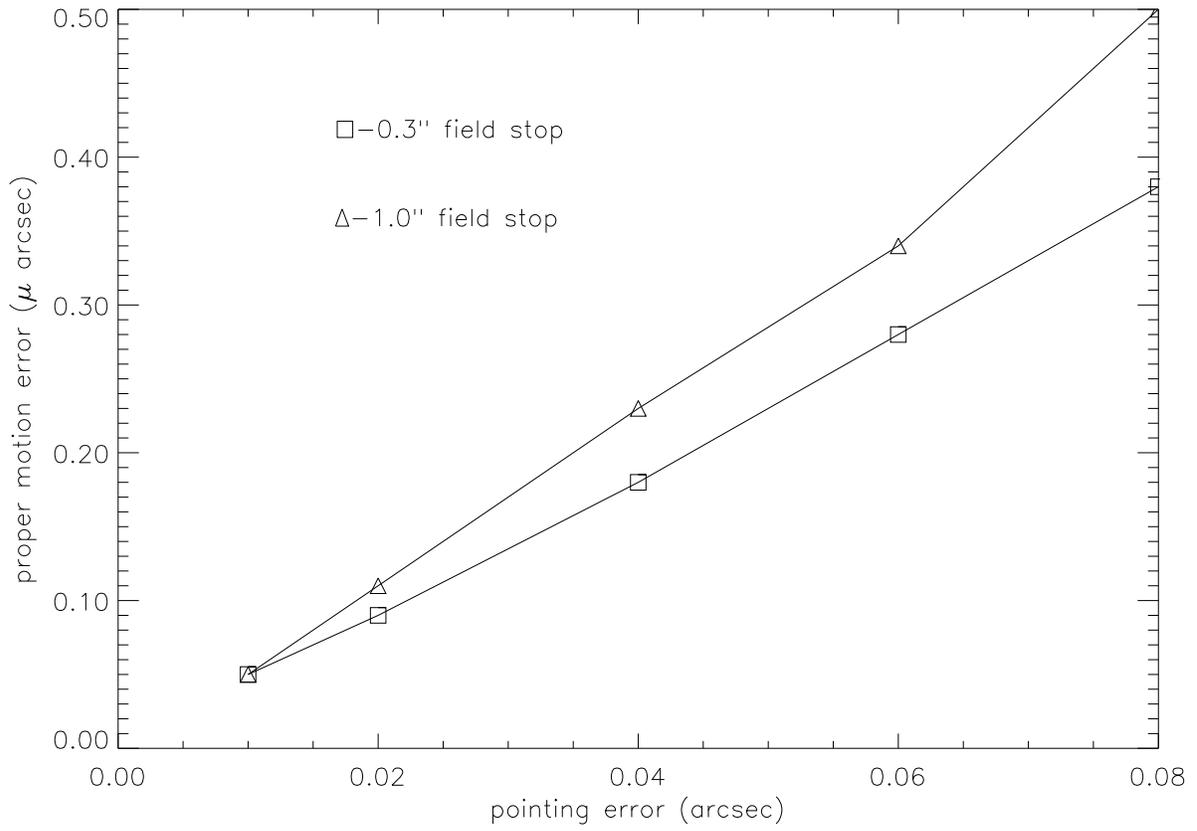}
\figcaption[f5.ps]{Proper motion error for the M31 field, with
parameters the same as in Figure 4 except for the bandwidth which is now $12\>{\rm nm}$. 
\label{fig-5}}
\end{figure}
\begin{figure}
\plotone{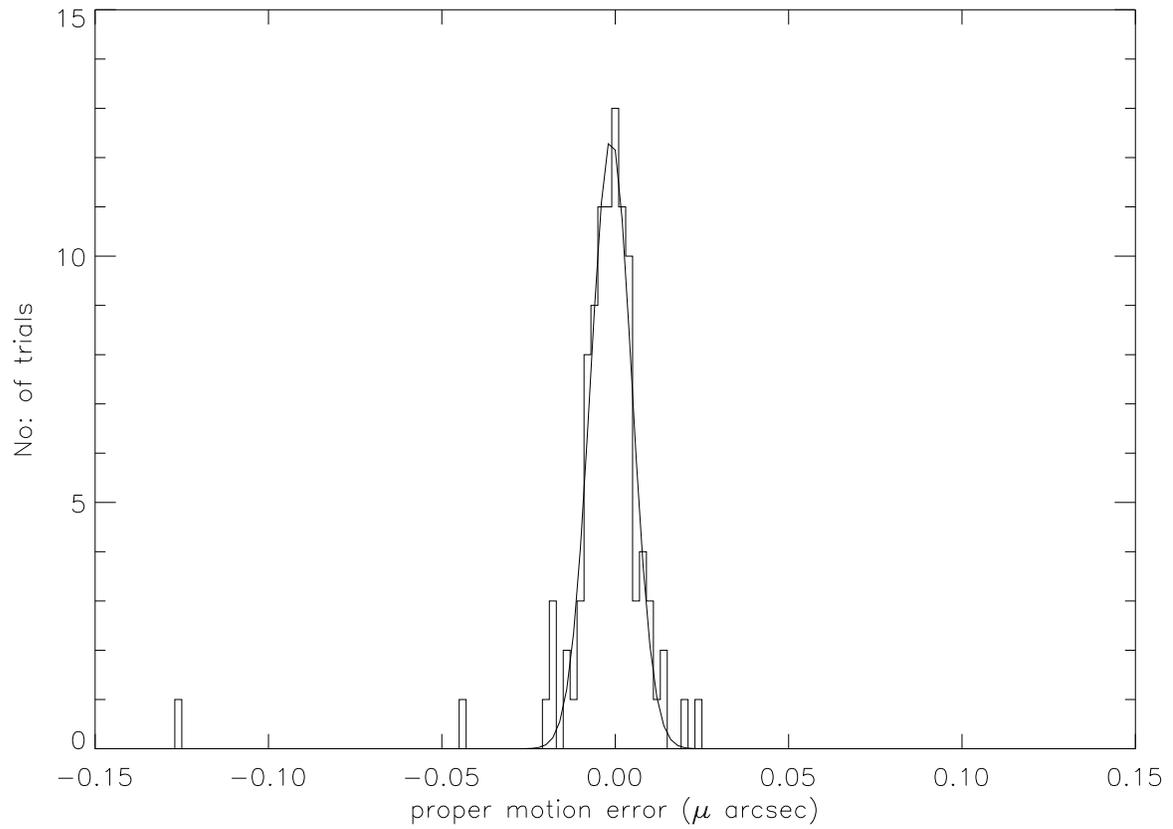}
\figcaption[f6.ps]{Histogram of the error in proper motion for the M31 field for
100 trials using
a pointing offset of 0.01\arcsec , with other parameters the same as in Figure 4. The width
of the distribution is a measure of the proper motion error. The Gaussian fit to the 
central peak is shown.\label{fig-6}} 
\end{figure}
\clearpage
\begin{figure}
\plotone{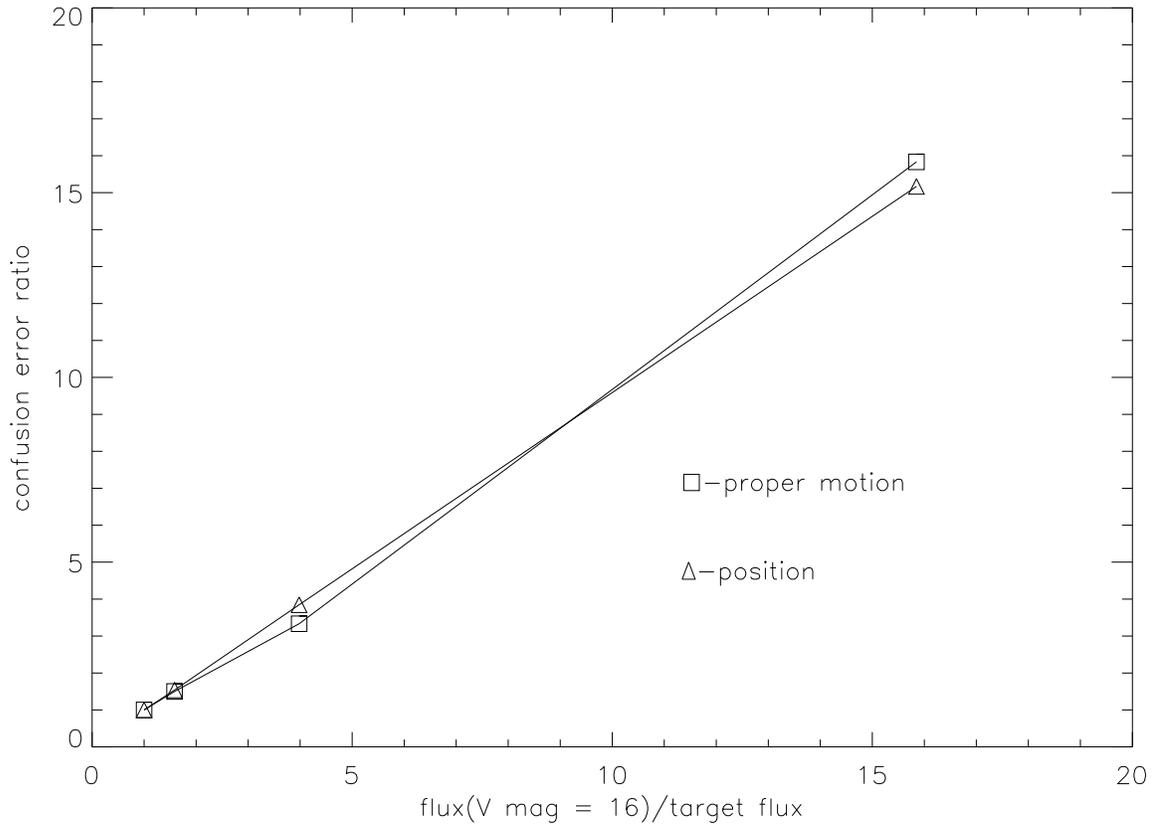}
\figcaption[f7.ps]{Scaling of confusion errors with target flux (broadband). The squares 
show the proper motion error ratio (normalised to that for a target with ${\rm m_v = 16}$) and 
the triangles are for position error.
Both errors scale inversely as the flux. The bandwidth used is $400 \>{\rm nm}$ and field stop 
size is 1\arcsec.\label{fig-7}}
\end{figure}
\clearpage
\begin{figure}
\plotone{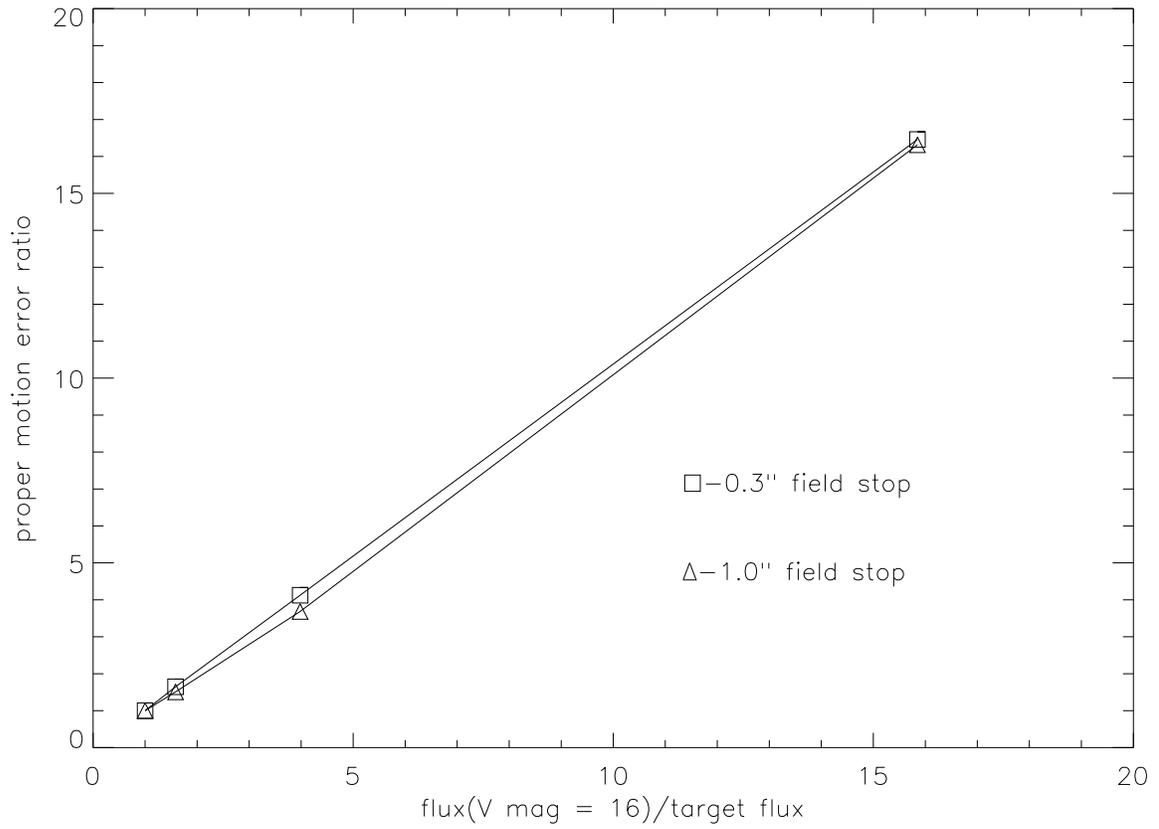}
\figcaption[mu_mag_12.ps]{Scaling of confusion errors with target flux (narrowband). The triangles 
show the proper motion error for a bandwidth of $12\>{\rm nm}$ and field stop size of 1\arcsec. The squares
are for a stop size of 0.3\arcsec.\label{fig-8}} 
\end{figure}
\clearpage
\begin{figure}
\plotone{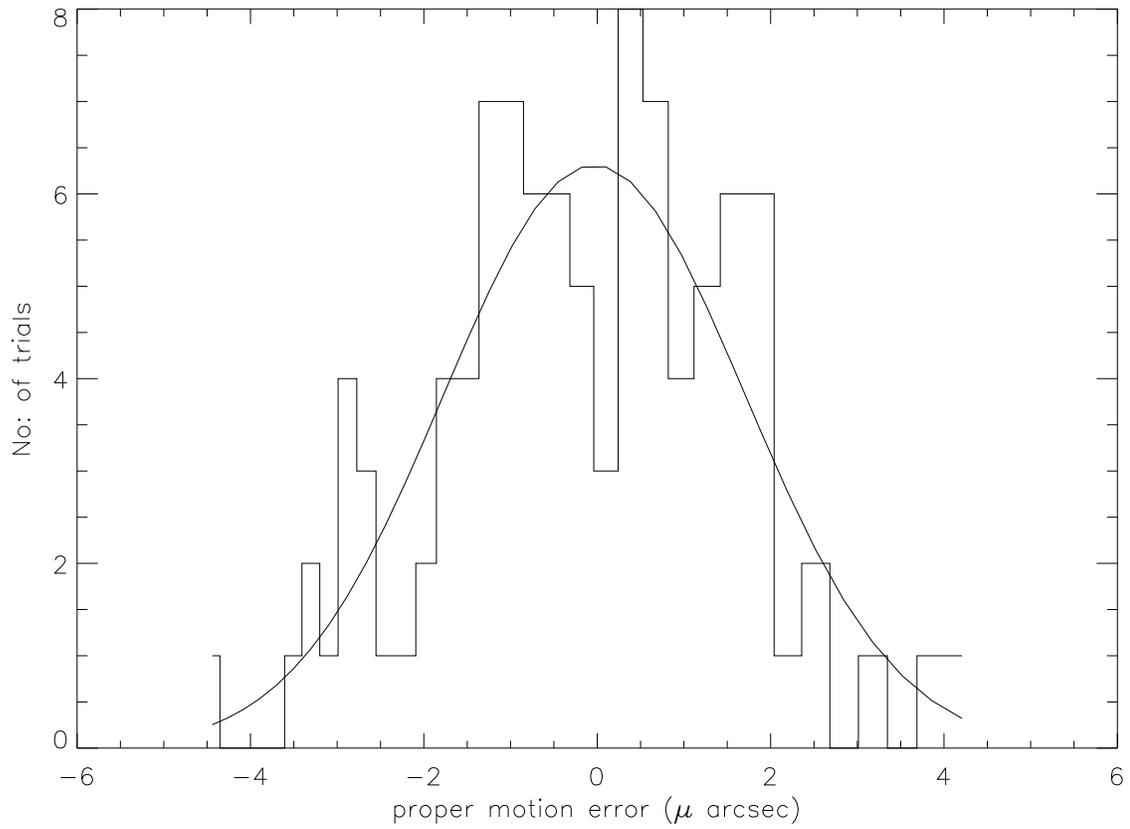}
\figcaption[f9.ps]{Histogram of proper motion errors with photon noise included. All
other parameters are the same as in Figure 4. The gaussian fit shown has a sigma of 2.0 $\mu$as.
Clearly, photon noise dominates over confusion (compare with Figure 6) for this case. \label{fig-9}} 
\end{figure}
\clearpage
\begin{figure}
\plotone{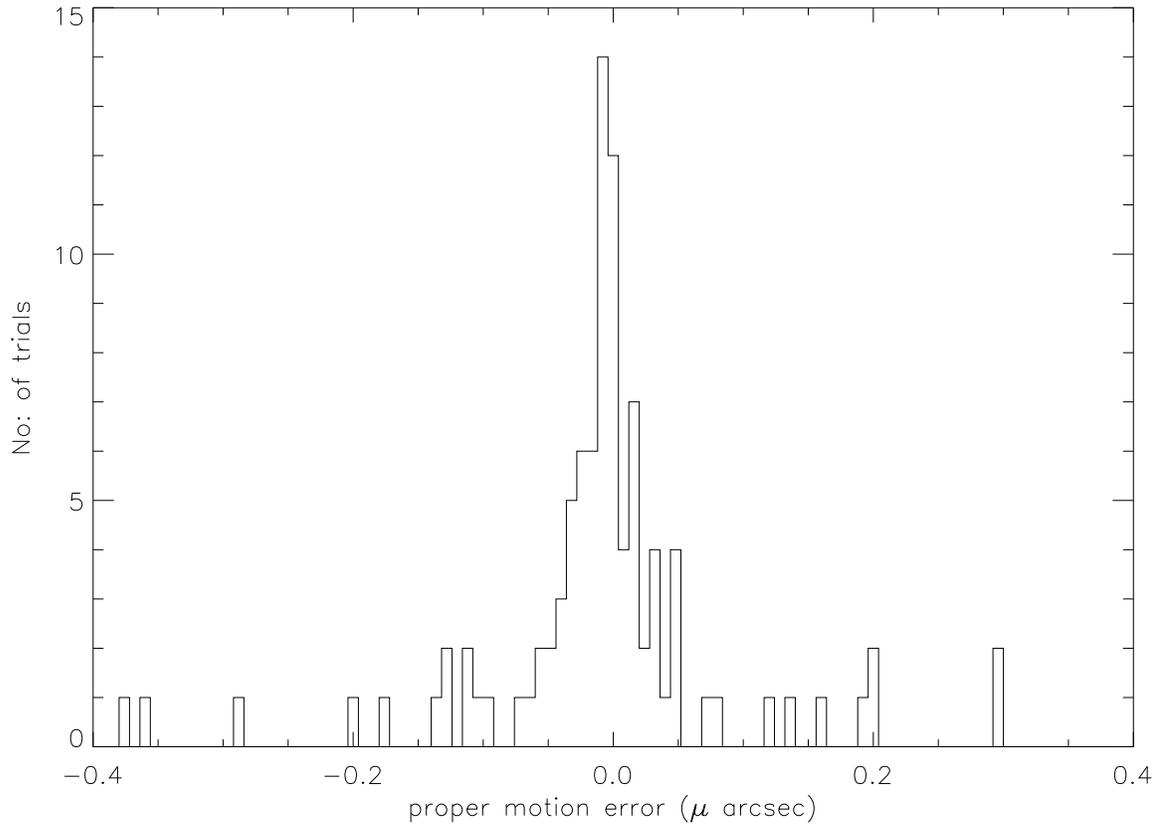}
\figcaption[f10.ps]{Histogram of proper motion error from confusion for the LMC field.
Compared to
Figure 6 (M31 field), the width of the distribution is narrow, indicating negligible confusion error.
The outliers occur when a comparatively bright source other than the target is present 
in  the FOV. \label{fig-10}}
\end{figure}
\clearpage
\begin{figure}
\plotone{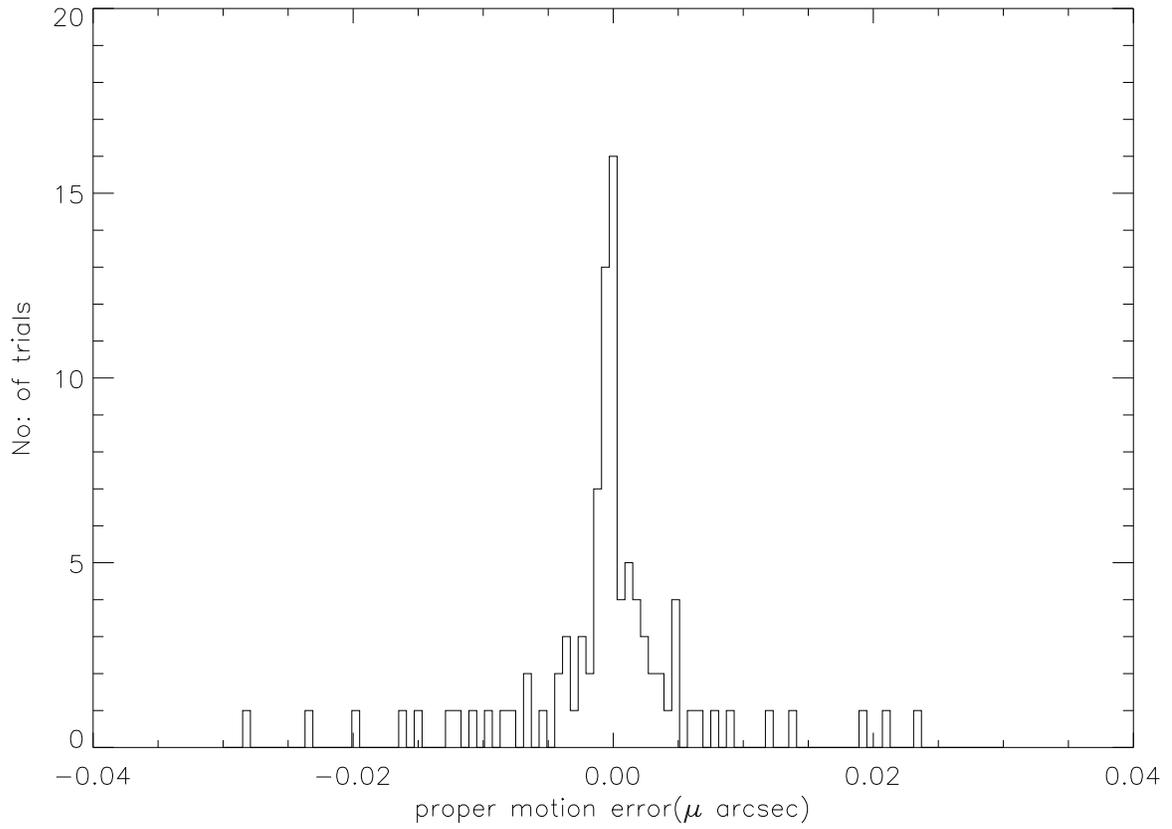}
\figcaption[f11.ps]{As in Figure 10, for a Galactic Bulge field. As in the LMC field, the distribution
is quite narrow except for a number of outliers. \label{fig-11}}
\end{figure}

\clearpage

\end{document}